\documentclass[preprint,aps,prb,showkeys,showpacs]{revtex4-1}

\usepackage{amssymb}
\usepackage{setspace}

\usepackage{color,graphicx,calc,url}
\usepackage{dcolumn}
\usepackage{bm}

\begin{document}

\title{Link between local iron coordination, Fe $2p$ XPS core level line shape and Mg segregation into thin magnetite films grown on MgO(001)}

\author{M. Suendorf$^1$, T. Schemme $^1$, A. Thomas$^2$, J. Wollschl\"ager$^1$, and K. Kuepper$^1$}
\email[]{kkuepper@uos.de}
\affiliation{	$^1$ Fachbereich Physik, Universit\"at Osnabr\"uck, Barbarastr. 7, 49069 Osnabr\"uck, Germany\\
							$^2$ Department of Physics, Bielefeld University, Universit\"atsstr. 25, D-33501 Bielefeld, Germany}				
					

\begin{abstract}

A well ordered Fe(001) ultra thin film epitaxially grown on MgO(001) has been oxidized by post deposition annealing in oxygen atmosphere. LEED patterns indicate the formation of magnetite (Fe$_3$O$_4$) after one hour of oxygen exposure. The LEED pattern remains stable despite annealing the sample for further four hours. In contrast X-ray Photoelectron Spectroscopy (XPS) of the Fe $2p$ core levels suggests the formation of an iron oxide comprising mostly of merely trivalent iron. This discrepancy is analyzed and discussed employing charge transfer multiplet calculations for the Fe $2p$ XPS spectra. We find that Mg ion segregation from the substrate into the magnetite thin film replacing octahedrally coordinated Fe$^{2+}$ ions and the excess occupation of octahedral sites which are unoccupied in the ideal inverse spinel structure are driving forces for the altered shape of the Fe $2p$ XPS spectra. Different potential models which might explain the nature of the Mg ion segregation into the magnetite films are discussed.

\end{abstract}


\maketitle

\section{Introduction}\label{Introduction}

Among a huge number of transition metal oxides, which display a remarkable variety of unusual transport properties and collective ordering phenomena\cite{mae04}, iron oxides are of particular interest. The importance of these materials stems partly from the quite different chemical and electronic properties in dependence of the Fe valence state, making iron oxides possible candidates for catalytic applications,\cite{ped99} and partly from the half metallic, ferromagnetic ground state of magnetite (Fe$_3$O$_4$)\cite{har96,zut04}. Magnetite crystallizes in the cubic inverse spinel structure structure (equal distribution of Fe$^{3+}$ on A and B sites and Fe$^{2+}$ exclusively on B sites) with lattice constant $a$=0.8396nm (space group Fd3m). The oxygen anions form an fcc anion sublattice. Due to the high Curie temperature of 584$^{\circ}$C, in particular Fe$_3$O$_4$ thin films are very interesting candidates for various potential future applications in magnetic data storage technology or for so called spintronic devices \cite{li98,fon05}. \par

\begin{figure}
	\includegraphics[width=7cm]{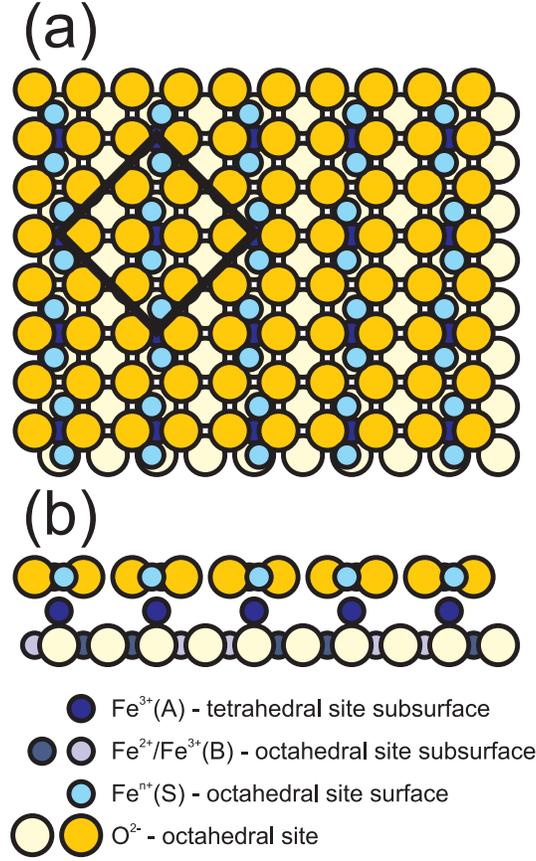}
	\caption{Model for the $(\sqrt{2}\times\sqrt{2})\rm{R}45^{\circ}$ reconstructed magnetite surface (a) top view and (b) side view. For the first subsurface B sublattice the Fe$^{2+}$ and Fe$^{3+}$ ions are charge ordered and the Fe ions of the B terminated surface form a wavy pattern. In addition, the tetrahedrally coordinated Fe$^{3+}$ of the A lattice occupy sites where adjacent octahedral sites of the B lattice are not occupied.
	\label{fig:InvSpinel}}
\end{figure}

The exact knowledge of the nature of the Fe$_3$O$_4$ surfaces and interfaces between magnetite and substrate are essential for the design of potential tunnel magneto resistance (TMR) building blocks \cite{li98,zha04}. As to the surface structure of magnetite, several models were presented for both A terminated and B terminated surfaces.
For instance, an early model presented by Chambers et al.\ assumed that the magnetite surface is A terminated and half of the regular A sites which are usually occupied by Fe$^{3+}$ ions are not occupied \cite{cha00}. Therefore, this model explained in a quite "natural way" the $(\sqrt{2}\times\sqrt{2})\rm{R}45^{\circ}$ reconstruction well known for magnetite.
In the meantime, however, more sophisticated models have been developed on the basis of calculations Density Functional Theory (DFT) \cite{pen05,lod07,pen08,mul09,mul10}. These models have been verified by several experiments including Grazing Incidence X-ray Diffraction (GIXRD) and Scanning Tunneling Microscopy (STM) \cite{pen05,par11a,par11b}.
The main ingredients of this model are, firstly, charge ordering in B subsurface layer inducing alternating Fe$^{2+}$ and Fe$^{3+}$ dimers forming a $(\sqrt{2}\times\sqrt{2})\rm{R}45^{\circ}$ pattern (cf. Fig.\ref{fig:InvSpinel}).
This charge ordering is, secondly, accompanied by a Jahn-Teller distorted anti-phase ordering of the Fe surface ions.
In addition, the Fe surface layers is Fe$^{3+}$ rich as reported from previous experiments with X-ray Photoelectron Spectroscopy (XPS) and X-ray Photoelectron Diffraction (XRD) \cite{cha00}.\par

MgO (cubic rock salt structure, a=0.42117nm) is one of the most promising substrates for epitaxial thin film growth of magnetite due to the rather small lattice mismatch of only 0.3\%, comparing the oxygen sublattices. Also, magnesia is commonly used as the barrier material in magnetic tunnel junctions, which are the building blocks for many spintronic applications \cite{dre09}.
However, potential Mg segregation into the grown Fe$_3$O$_4$ thin film is a major potential drawback, which usually occurs if the MgO substrate is heated to 300$^{\circ}$C and above during magnetite growth. Mg interdiffusion\cite{the00,kim08,kim09} and the formation of an MgO rich interface layer have been reported ´\cite{sha00,han01,kor02}. MgO segregation across the MgO/Fe$_3$O$_4$ interface is crucial since the influence of grain boundaries, interface roughness and in particular anti phase boundaries, which might partly stem from Mg segregation into the Fe$_3$O$_4$ layer, on the low-field magneto resistance response is at least partly controversial \cite{zie02,lie03,aro05,jin05,sof07,wu12}. In particular, Kalev et al.\ find the Fe ions at the Fe$_3$O$_4$/MgO interface to be in the Fe$^{3+}$ valence state\cite{kal03}, leading to a significant reduction of spin fluctuations and magnetic neighbors at the interface. Very recently it has been found that additional electron scattering at the anti-phase boundaries has to be considered in order to explain the transversal magneto resistance in epitaxial MgO/Fe$_3$O$_4$ thin film systems\cite{wu12}. Hence, a detailed characterization of the interface including the presence of antiphase boundaries, width of structural domains and other crystallographic effects which may be introduced by Mg ion segregation or diffusion during the growth process is an indispensable tool if one wants to understand the contradictory results according to the magnetotransport properties of MgO/Fe$_3$O$_4$ multilayers.\par

The modification of the electronic, magnetic and structural properties due to Mg$^{2+}$ ion segregation and diffusion into the magnetite layer has been mostly associated with the replacement of octahedrally coordinated Fe$^{2+}$ ions with Mg$^{2+}$ ions\cite{and97}. Here we go a step further and investigate the surface and interface of MgO/Fe$_3$O$_4$ bilayers grown by molecular beam epitaxy of Fe with subsequent post deposition annealing oxidation by means of low energy electron diffraction (LEED) and x-ray photoelectron spectroscopy (XPS). The Fe $2p$ core level spectra are furthermore analyzed by means of charge transfer multiplet (CTM) simulations. Mg$^{2+}$ segregation onto octahedrally coordinated Fe$^{2+}$ sites is one process taking place during the Mg interdiffusion process into the magnetite layer. Furthermore, additional analysis of the Mg:Fe ratio suggests that Mg$^{2+}$ ions also occupy parts of the unoccupied octahedral sites of the ideal inverse spinel structure.

\section{Experimental setup and sample preparation}\label{Experiment}

The experiments were performed in a multi-chamber ultra high vacuum (UHV) system with a base pressure of $10^{-10}\,\rm{mbar}$.
The XPS system consisted of a SPECS XR50 x-ray source providing Al K$_{\alpha_{1,2}}$ radiation at $1486.6\,\rm{eV}$ and a PHOIBOS 150 hemispherical electron analyzer with a resolution of $1\,\rm{eV}$.
LEED images were obtained with ErLEED 150 optics.

Polished MgO(001) substrates (miscut $< 1^{\circ}$) were cleaned {\it in-situ} by annealing for one hour at $700^{\circ}\rm{C}$ in $10^{-4}\,\rm{mbar}$ oxygen atmosphere.
Afterwards purity and quality of the substrate surface were checked by XPS and LEED prior to film deposition.

An iron oxide film was prepared in two steps.
First, a pure iron film was deposited by molecular beam epitaxy (MBE) from an iron rod heated by electron bombardment.
The amount of deposited material was monitored by a balance quartz.
The MgO(001) substrate was held at room temperature during deposition.
XPS showed that the Mg $2p$ signal from the substrate was completely suppressed, thus the deposited film was at least $5\,\rm{nm}$ thick.
Also careful analysis of the Fe 2p spectrum showed no signal of iron oxide was observed at this stage, indicating a pure metallic iron film.

The iron film was then treated in $10^{-6}\,\rm{mbar}$ oxygen atmosphere at a temperature of 300$^{\circ}\rm{C}$.
This process was performed in cycles of one hour of oxygen exposure followed by XPS and LEED measurements.
Because of the insulating character of the substrate strong charging effects were observed in the photoemission (PE) spectra.
Therefore, the recorded PE spectra were calibrated to the binding energy (BE) of the O $1s$ signal at $530\,\rm{eV}$.

\section{Calculation method}\label{Calculation}

A charge transfer multiplet (CTM) model were used to calculate the Fe $2p$ PE spectra of iron oxides.
The approach is based on a cluster model.
It is especially useful to describe core level spectra of transition metals because of the strong influence of unpaired electrons in the $d$ shells\cite{fuj99}.
Within this method the sample is described by clusters consisting of a single metal ion in the center and the influence of symmetrically aligned neighboring oxygen ligands.
The ground state of the metal ion is assumed to be a linear combination of the regular electronic configuration $3d^n$ and the charge transfer configurations $3d^{n+1}\underline L$ and $3d^{n+2}\underline L^2$ where one and two electrons have been transferred from the oxygen ligand shell $L$ to the metal ion, respectively.
$n$ denotes the number of $d$ electrons of the metal ion and $\underline L$ denotes a hole in the oxygen ligand shell.
Similarly, the final state after emission of the Fe $2p$ photo electron is described by a linear combination of the configurations $\underline c 3d^n$, $\underline c 3d^{n+1}\underline L$ and $\underline c 3d^{n+2}\underline L^2$, where $\underline c$ denotes a hole in a metal core shell.
The energy difference between the configurations is defined via the Coulomb interaction $Q$ between core hole and $d$ electron.
Compared to the normal configuration, the two charge transfer states are lowered in energy by $Q$ and $2Q$, respectively.
It has been reported that this reordering of states is the reason for the strong satellites in the PE spectra of transition metals\cite{zaa85}.

In the CTM calculation in addition to the Coulomb interaction the correlation energy ($U$) between two $d$ electrons, the oxygen $2p$ to metal $3d$ charge transfer energy ($\Delta$) and the oxygen $2p$ to metal $3d$ hybridization energy ($T$) are explicitly taken into account.
Correlations between the parameters and a set of values able to describe experimental data of Fe $2p$ PE spectra were discussed previously \cite{fuj99}.
We therefore limit our analysis to the literature values of these parameters as presented in Table \ref{tab:CTM_parameters}.
Following Fuji {\it et al.}\cite{fuj99} the crystal field strength was fixed at $1\,\rm{eV}$ for octahedral symmetry because this parameter had only minor influence on the spectral shape.
In tetrahedral symmetry the sign of the crystal field is reversed.


\begin{table}[ht]
	\begin{tabular}{c|c|c|c|c|c}
	Fe site & $\Delta$ ($\rm{eV}$) & T(e$_g$) ($\rm{eV}$) & T(t$_{2_g}$) ($\rm{eV}$) & U ($\rm{eV}$) & Q ($\rm{eV}$) \\ \hline
	Fe$^{2+}_{oct}$ & $4{.}0$ & $2{.}3$ & $-1{.}15$ & $7{.}0$ & $7{.}5$ \\
	Fe$^{3+}_{oct}$ & $2{.}0$ & $2{.}2$ & $-1{.}1$ & $7{.}5$ & $8{.}0$ \\
	Fe$^{3+}_{tet}$ & $2{.}0$ & $1{.}35$ & $-2{.}7$ & $7{.}0$ & $7{.}5$ \\
	\end{tabular}
	\caption{Parameter values used for the cluster calculations of the Fe $2p$ PE spectrum of Fe$_3$O$_4$. Taken from\cite{fuj99}.}\label{tab:CTM_parameters}
\end{table}

\section{Results}\label{Results}

\subsection{Characterization of the pure Fe film}

\begin{figure}
	\includegraphics[width=9.5cm]{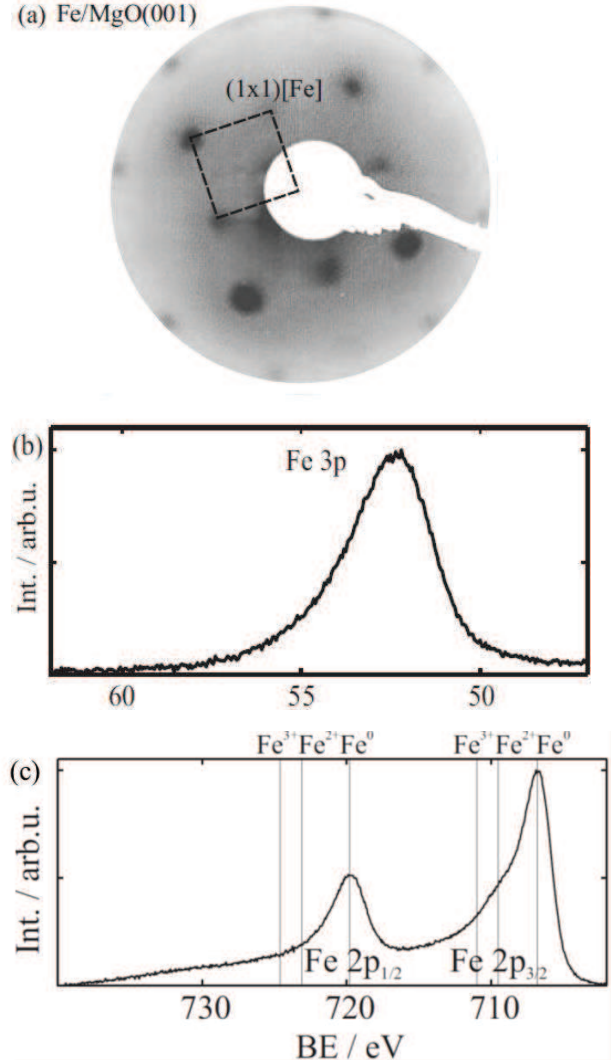}
	\caption{(a) LEED image of an Fe film grown on MgO(001). The $(1\times 1)$ Fe surface unit cell is indicated. (b) and (c) PE spectra of the Fe $3p$ and $2p$ regions of the same Fe film.\label{fig:Fe}}
\end{figure}

LEED and XPS results of the pure Fe film after deposition are shown in Fig.\ \ref{fig:Fe}.
The film exhibits the clear $(1\times 1)$ LEED pattern of Fe(001) presented in Fig.\ \ref{fig:Fe}(a).
No additional diffraction spots are observed.
Compared to the LEED pattern obtained from the MgO(001) substrate, however, the diffraction spots obtained from the Fe(001) film are broadened.
This points to some disorder for the RT grown Fe film.
The PE spectrum of the Fe $3p$ region in Fig.\ \ref{fig:Fe} (b) shows that the Mg $2p$ signal at $49\,\rm{eV}$ is completely suppressed, indicating a closed film of at least $5\,\rm{nm}$ thickness.
The Fe $2p$ peak in Fig.\ \ref{fig:Fe} indicates that the film consists mainly of metallic iron (cf.\ Fig.\ \ref{fig:Fe} (c)). The position of the main peak corresponds well to literature data for metallic iron. However, the shoulder located at around 709.5 eV (literature data for Fe$^{2+}$ and Fe$^{3+}$ are indicated by vertical bars) indicates the presence of some divalent iron oxide at the surface of the film.

\subsection{Qualitative analysis of the oxidation process}

\begin{figure}
	\includegraphics{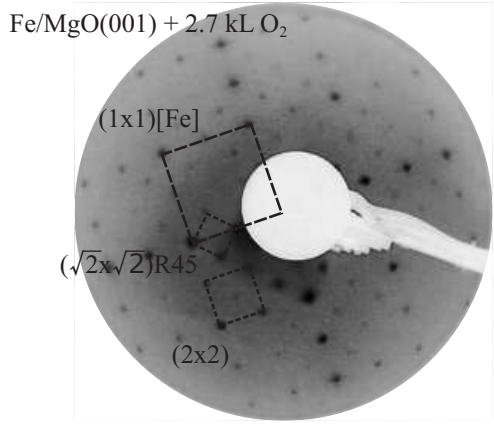}
	\caption{LEED pattern of Fe/MgO(001) after one hour in oxygen atmosphere at $300^{\circ}\rm{C}$. A $(\sqrt{2}\times\sqrt{2})\rm{R}45^{\circ}$ reconstruction with respect to the Fe surface unit cell is indicated.\label{fig:LEED}}
\end{figure}

Fig.\ \ref{fig:LEED} shows the LEED pattern of the film after one hour of exposure to oxygen.
Compared to the initial Fe film additional spots can be observed.
The diffraction pattern proves that the surface lattice constant has approximately doubled and, in addition, a $(\sqrt{2}\times\sqrt{2})\rm{R}45^{\circ}$ reconstruction has formed.
The appearance of this reconstruction shows that the film structure has changed from the bcc lattice of $\alpha$-Fe to a more complex structure with a larger unit cell size.
In particular, the observed reconstruction is usually associated with the formation of magnetite (Fe$_3$O$_4$) \cite{pen05,lod07,pen08,mul09,mul10}.
This LEED pattern does not change visibly with prolonged oxidation.

\begin{figure}
	\includegraphics{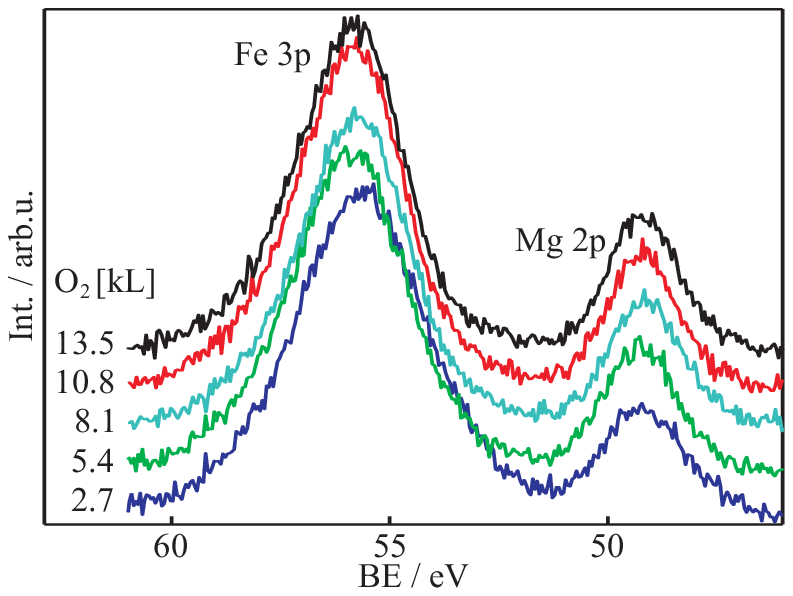}
	\caption{Fe $3p$ PE spectra for each step of oxygen exposure (one hour in $10^{-6}\,\rm{mbar}$ oxygen atmosphere at 300$^{\circ}\rm{C}$).\label{fig:XPS_3p}}
\end{figure}

The PE spectra of the Fe $3p$ region during oxygen exposure are given in Figure \ref{fig:XPS_3p}.
The spectra were taken at one hour intervals during oxidation (equivalent to exposure intervalls of 2700 L).
The Fe $3p$ peak is shifted by $3{.}5\,\rm{eV}$ to a higher BE.
The new peak position corresponds well to known values for Fe oxides\cite{yam08}.
A signal of metallic Fe is not observed anymore.
However, a clear signal of Mg $2p$ at $49\,\rm{eV}$ reappears in the spectrum, too.
The ratio between Fe and Mg in the film is approximately $1:1$ and remains constant during the entire exposure to oxygen.
Thus the diffusion of Mg in the iron oxide film is already completed after the first oxidation step.
In addition, the O $1s$ photoemission intensity has also been analyzed (not shown here).
Compared to the cation PE intensities the O $1s$ intensity is constant.
Finally, we can conclude that the oxide film has the Fe:Mg:O stoichiometry of 1:1:3.
These observations show that the film is completely oxidized to a depth of at least a few $\rm{nm}$.
Changes in the spectral shape of the Fe $3p$ peak can not be observed for prolonged oxidation.

\begin{figure}
	\includegraphics{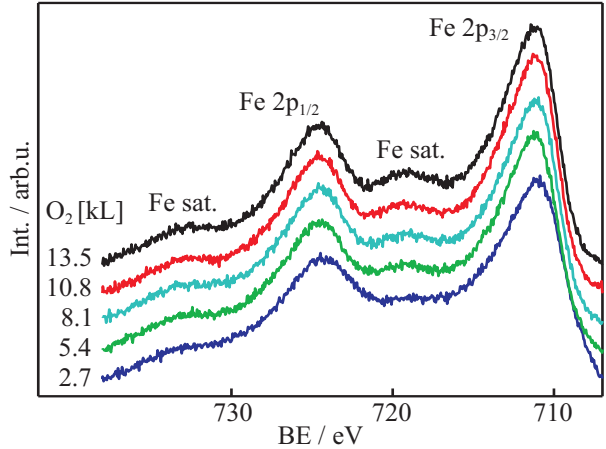}
	\caption{Fe $2p$ PE spectra for each step of oxygen exposure (one hour in $10^{-6}\,\rm{mbar}$ oxygen atmosphere at 300$^{\circ}\rm{C}$).\label{fig:XPS_2p}}
\end{figure}

Figure \ref{fig:XPS_2p} shows the PE spectra of the Fe $2p$ region during oxidation.
The same steps of oxidation as in Figure \ref{fig:XPS_3p} are given.
Again a chemical shift from metallic Fe to Fe oxide is apparent, after the first oxidation step no metallic Fe is detected for the oxidized film.
After one hour of exposure to oxygen a plateau of constant intensity between the main Fe $2p_{3/2}$ and Fe $2p_{1/2}$ peaks is observed.
This spectral shape is usually associated with the formation of magnetite (Fe$_3$O$_4$) as shown in Figure \ref{fig:XPS_2p}, too \cite{cha99}.
With prolonged oxidation a strong satellite peak at about $8\,\rm{eV}$ higher BE from the Fe $2p_{3/2}$ peak gradually appears in the recorded PE spectra.
This satellite is usually attributed to a stoichiometry involving a higher Fe$^{3+}$ content which, without any Fe$^{2+}$ content, may finally form maghemite\cite{cha99} ($\gamma$-Fe$_2$O$_3$).

\subsection{Quantitative analysis using CTM calculations}

The observed LEED patterns demonstrate that all prepared Fe oxide films have a surface superstructure that is typical for the magnetite Fe$_3$O$_4$(001) surface.
This structure is characterized by three different Fe ion sites in the crystal lattice: Fe$^{2+}$ and Fe$^{3+}$ on octahedrally coordinated sites and Fe$^{3+}$ on tetrahedrally coordinated sites.
Although the $(\sqrt{2}\times\sqrt{2})\rm{R}45^{\circ}$ LEED pattern of the film remains unchanged regardless of oxygen exposure, PE spectra of the Fe $3p$ region show that the stoichiometry of the film is not pure iron oxide but the oxide is mixed with Mg.
Furthermore, the Fe $2p$ spectra indicate considerable changes of the Fe$^{2+}$-Fe$^{3+}$ distribution during post deposition oxidation.
A careful analysis with the help of CTM calculations has to be performed to explain these discrepancies.

\begin{figure}
	\includegraphics{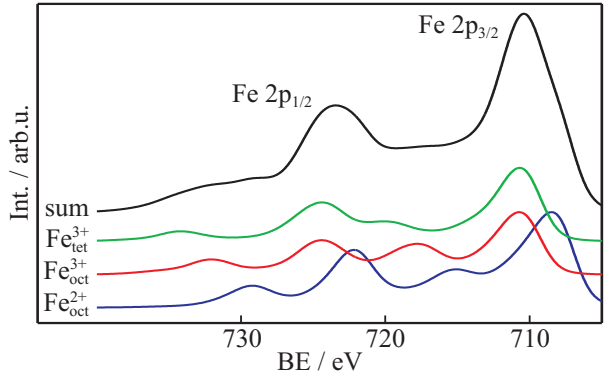}
	\caption{Charge Transfer Multiplet calculations for the Fe sites in Fe$_3$O$_4$. From bottom to top: Fe$^{2+}_{\rm{oct}}$, Fe$^{3+}_{\rm{oct}}$ and Fe$^{3+}_{\rm{tet}}$. The resulting stoichiometric summation gives the expected magnetite spectrum (top).\label{fig:cluster}}
\end{figure}

CTM theory allows the calculation of PE spectra for each metal site in a crystal lattice individually.
This is demonstrated in Figure \ref{fig:cluster}.
In stoichiometric Fe$_3$O$_4$ the three metal ion sites Fe$^{2+}_{\rm{oct}}$, Fe$^{3+}_{\rm{oct}}$ and Fe$^{3+}_{\rm{tet}}$ should be occupied equally.
All three CTM spectra show distinct satellites with apparent higher BE than the main Fe $2p$ peaks.
For the Fe$^{2+}$ octahedral site these satellites are located closest to the fundamental peaks ($\Delta E\approx 6{.}5\,\rm{eV}$).
The energy separation is slightly larger for trivalent octahedral sites ($\Delta E\approx 7\,\rm{eV}$) and is largest for the tetrahedral site ($\Delta E\approx 9\,\rm{eV}$).
A stoichiometric summation of the three CTM spectra results in the expected magnetite spectrum without any distinct satellite due to superposition.
However, if one incorporates Fe vacancies in the oxide film, one or more of the sites will have a decreased contribution to the total spectrum and their superposition leads to drastically altered spectra.
We will now demonstrate that this is the reason for the observed changes in the Fe $2p$ spectra presented in Figure \ref{fig:XPS_2p}.

We start our analysis on the common assumption that Mg$^{2+}$ ions gradually replace some of the Fe$^{2+}$ ions on octahedral lattice sites in the film to keep the charge neutrality of the structure.
Then the contribution of the respective Fe site in the superposition of the CTM spectra is decreased depending on the ratio of substituted ions.
The individual CTM calculations themselves are not influenced because they are independent of neighboring metal sites and only depend on the oxygen ion environment.

\begin{figure}
	\includegraphics{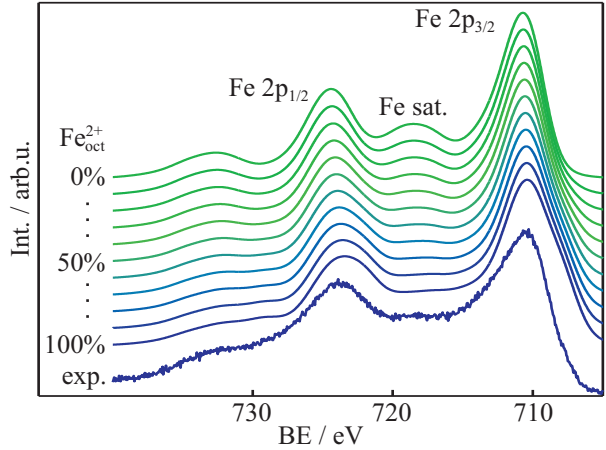}
	\caption{CTM calculation of the Fe $2p$ region for a Fe$_3$O$_4$ structure. The ratio of Fe ions occupying Fe$^{2+}_{\rm{oct}}$ sites is gradually decreased from $100\%$ to $0\%$ (from bottom to top) while Fe$^{3+}_{\rm{oct}}$ and Fe$^{3+}_{\rm{tet}}$ are held fixed at $100\%$.\label{fig:CTM_1o_Bsp1}}
\end{figure}

Fig.\ \ref{fig:CTM_1o_Bsp1} shows calculations where the ratio of Fe ions occupying Fe$^{2+}_{\rm{oct}}$ sites is gradually decreased from $100\%$ to $0\%$ while the occupation of Fe$^{3+}_{\rm{oct}}$ and Fe$^{3+}_{\rm{tet}}$ sites is held constant at 100\%.
For comparison the experimental Fe $2p$ spectrum after one hour of oxygen exposure is shown, too.
Inspection by eye already shows immediately that Fe$^{2+}$ ions can not be completely replaced by Mg$^{2+}$ ions since this would result in the spectrum at the top (0\% Fe$^{2+}$ content), which shows a pronounced satellite between the Fe $2p_{1/2}$ and the Fe $2p_{3/2}$ peaks.
Instead one can conclude that the Fe$^{2+}$ content is approximately 40\% due to the weak apparent satellite at 718 eV.
We have also to conclude that there must be many Mg$^{2+}$ ions on non-regular sites of the inverse spinell because that the Mg:Fe ratio is in the order of $1:1$ as concluded from the Fe $3p$ and Mg $2p$ spectra.

\begin{figure}
	\includegraphics{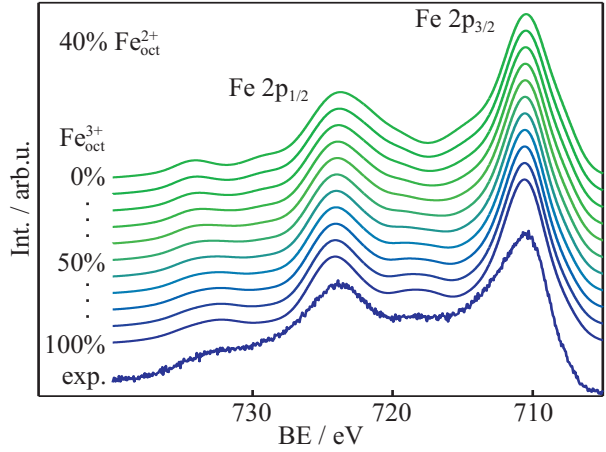}
	\caption{CTM calculation of the Fe $2p$ region for a Fe$_3$O$_4$ structure. The ratio of Fe ions occupying Fe$^{3+}_{\rm{oct}}$ sites is gradually decreased from $100\%$ to $0\%$ (from bottom to top). The ratio of Fe on Fe$^{2+}_{\rm{oct}}$ sites is fixed at $40\%$, Fe$^{3+}_{\rm{tet}}$ is held constant at $100\%$.\label{fig:CTM_1o_Bsp2}}
\end{figure}

In the next step we therefore assume that Mg also influences the Fe occupation of other lattice sites as well. To be more specific, we assume that all occupied octahedral and tetrahedral sites may be influenced by the interdiffusion of magnesium.
Figure \ref{fig:CTM_1o_Bsp2} demonstrates results for varying ratios of Fe occupying Fe$^{3+}_{\rm{oct}}$ sites while the occupation of the Fe$^{2+}_{\rm{oct}}$ and the Fe$^{3+}_{\rm{tet}}$ sites was fixed to $40\%$ and 100\%, respectively.
We characterize the quality of the calculated spectra by comparing position and intensity of the calculated Fe satellite to the experimental data.
Obviously many unphysical combinations of CTM spectra yield unrealistic spectral shapes, especially if the ratio of Fe on Fe$^{3+}_{\rm{tet}}$ sites is also allowed to vary.

\begin{figure}
	\includegraphics{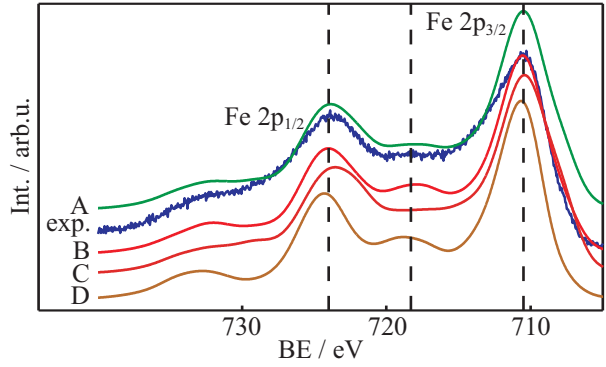}
	\caption{Agreement between CTM calculations and the measured Fe $2p$ PE spectrum after one hour of oxygen exposure. The ratios used in these calculations are given in Table \ref{tab:CTM_1o}.\label{fig:CTM_1o}}
\end{figure}

Following these aspects we obtain percentage regions that result in good agreement between calculation and experiment.
This is shown in Figure \ref{fig:CTM_1o} where different exemplary calculated spectra are compared to the experimental spectrum obtained after one hour of oxygen exposure.
A good agreement (cf. Figure \ref{fig:CTM_1o} A) is only achieved if the (normalized) ratio between Fe ions occupying Fe$^{2+}_{\rm{oct}}$, Fe$^{3+}_{\rm{oct}}$ and Fe$^{3+}_{\rm{tet}}$ lattice sites is $0{.}24:0{.}38:0{.}38$.
Other ratios result in strongly shifted main or satellite peaks or wrong satellite intensities.
The ratios used in these exemplary examples are given in Table \ref{tab:CTM_1o}.
It can be seen that at this stage of oxygen exposure Mg$^{2+}$ ions prefer to diffuse to Fe$^{2+}_{\rm{oct}}$ sites while Fe$^{3+}_{\rm{oct}}$ and Fe$^{3+}_{\rm{tet}}$ sites are still mostly occupied by Fe ions.



\begin{table}
	\begin{tabular}{c|c|c|c|c}
	& Fe$^{2+}_{\rm{oct}}$ & Fe$^{3+}_{\rm{oct}}$ & Fe$^{3+}_{\rm{tet}}$ & fit\\ \hline
	A & $0{.}24$ & $0{.}38$ & $0{.}38$ & good\\
	B & $0{.}2$ & $0{.}47$ & $0{.}33$ & bad\\
	C & $0{.}32$ & $0{.}32$ & $0{.}36$ & bad\\
	D & $0{.}07$ & $0{.}4$ &$0{.}53$ & bad\\
	\end{tabular}
	\caption{(Normalized) ratios between Fe sites occupied by Fe ions for the CTM calculations in Figure \ref{fig:CTM_1o}. The first row gives a good fit between calculation and experiment, the last three rows give wrong spectral shapes.\label{tab:CTM_1o}}	
\end{table}

\begin{figure}
	\includegraphics{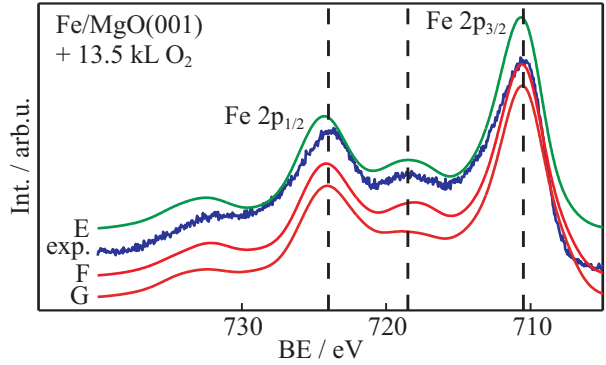}
	\caption{Agreement between CTM calculations and the measured Fe $2p$ PE spectrum after five hours of oxygen exposure. The ratios used in these calculations are given in Table \ref{tab:CTM_5o}.\label{fig:CTM_5o}}
\end{figure}

A similar analysis of the spectrum obtained after five hours of post deposition oxidation is presented in Figure \ref{fig:CTM_5o}.
Exemplary calculated spectra for different Fe ion occupations of octahedral and tetrahedral sites are shown and compared with the measured spectrum.
The Fe ion ratios used in Figure \ref{fig:CTM_5o} are given in Table \ref{tab:CTM_5o}.
In this case best results are obtained for the normalized ratio between Fe on
Fe$^{2+}_{\rm{oct}}$, Fe$^{3+}_{\rm{oct}}$
and Fe$^{3+}_{\rm{tet}}$ lattice sites is $0{.}06:0{.}47:0{.}47$.
Obviously, the amount of Fe$^{2+}_{\rm{oct}}$ is substantially reduced while Fe$^{3+}_{\rm{oct}}$ and Fe$^{3+}_{\rm{tet}}$ lattice sites are still equally occupied.
Thus, a higher amount of Mg ions seems to accumulate on Fe$^{2+}_{\rm{oct}}$ sites with longer oxidation time.
This effect may be attributed to re-ordering between Mg ions and Fe ions and to additional cation sites that are be generated by continuing the oxidation process.
This will be discussed in more detail below.



\begin{table}
	\begin{tabular}{c|c|c|c|c}
	& Fe$^{2+}_{\rm{oct}}$ & Fe$^{3+}_{\rm{oct}}$ & Fe$^{3+}_{\rm{tet}}$ & fit\\ \hline
	E & $0{.}06$ & $0{.}47$ & $0{.}47$ & good\\
	F & $0{.}14$ & $0{.}5$ & $0{.}36$ & bad\\
	G & $0{.}18$ & $0{.}35$ & $0{.}47$ & bad\\
	\end{tabular}
	\caption{(Normalized) ratios between Fe sites occupied by Fe ions for the CTM calculations in Figure \ref{fig:CTM_5o}. The first row gives a good fit between calculation and experiment, the last two rows give wrong spectral shapes.\label{tab:CTM_5o}}
\end{table}

\section{Discussion}\label{Discussion}

The segregation of Mg$^{2+}$ ions from a MgO substrate into an iron oxide film is widely discussed in literature\cite{and97,gao97,spi04,kim09}.
Commonly, it is assumed that these Mg$^{2+}$ ions exclusively replace Fe$^{2+}$ ions which occupy only octahedral cation sites of the inverse spinel structure of magnetite and thus, creating a Mg XPS signal.
Further it is assumed that during the intermixing process the Fe$^{3+}$ ions stay on the octahedral and tetrahedal sites where they are located in the magnetite structure.
Thus, the stoichiometry of this Mg ferrite structure can be denoted by Mg$^{2+}_{\rm{x}}$Fe$^{2+}_{\rm{1-x}}$Fe$^{3+}_{\rm{oct}}$Fe$^{3+}_{\rm{tet}}$O$_4$ with maximum Mg:Fe ratio of 1:2.

Here, we present detailed XPS studies on the Fe $2p$, Fe $3p$ and Mg $2p$ PE spectra to follow the Mg ferrite formation during the oxidation process of Fe films on MgO(001) at elevated temperatures.
Information concerning the content of Fe and Mg in the oxide films are obtained from comparison of the Fe $3p$ and the Mg $2p$ PE signal while the distribution of Fe$^{n+}$ ions on different sublattice sites can be deduced from Fe $2p$ spectra.
For the latter purpose a detailed CTM analysis of the Fe $2p$ PE spectra has been performed.

The CTM analysis of the Fe $2p$ signal agrees with the formation of Mg ferrite since the contents of Fe$^{3+}_{\rm{oct}}$ and Fe$^{3+}_{\rm{tet}}$ are equal for each oxidation step while the ratio Fe$^{2+}$:Fe$^{3+}$ (Fe$^{3+}$ on both octahedral and tetrahedral sites) decreases continuously.
These results, however, describe only the relative Fe occupation on the different sublattices.
We like to emphasize that occupation of adjacent cation sites does not influence our CTM analysis since these calculations consider only the local oxygen ligand structure.
Any additional positive charges introduced into the unit cells here can be balanced by a reordering of the Fe cations.

While on the one hand the Fe $2p$ analysis points to ferrite formation, on the other hand, this analysis has to be modified considering the combined Fe $3p$-Mg $2p$ PE spectra.
They show that the Mg:Fe ratio is 1:1 which exceeds the maximum Mg:Fe ratio of 1:2 well-known for Mg ferrite.
Taking into account theses spectroscopic results and assuming that the oxygen anion sublattice retains its fcc symmetry without any formation of oxygen vacancies, the total stoichiometry of the high temperature oxidized Fe film can be described by Mg$^{2+}_{\rm{x}}$Fe$^{2+}_ {\rm{y}}$(Fe$^{3+}_{\rm{oct}}$Fe$^{3+}_{\rm{tet}}$)$_{\rm{z}}$O$_4$.
In detail, we obtain x$>$z and y$<$z for the films studied here after the different oxidation steps.

Thus, these oxidized Fe films contain excess Mg$^{2+}$ compared to the complete ferrite formation of Mg$^{2+}_{\rm{oct}}$Fe$^{3+}_{\rm{oct}}$Fe$^{3+}_{\rm{tet}}$O$_4$.
The excess of Mg$^{2+}$ ions is still larger if one takes into account that Mg$^{2+}$ ions can only replace Fe$^{2+}$ ions on octahedral sites and that for the first oxidation steps the content of Fe$^{2+}$ ions is still considerable.
Therefore, Mg$^{2+}$ ions have to occupy additional sites which are vacant in the inverse spinel structure of magnetite and ferrites.

\begin{figure}
	\includegraphics[width=7cm]{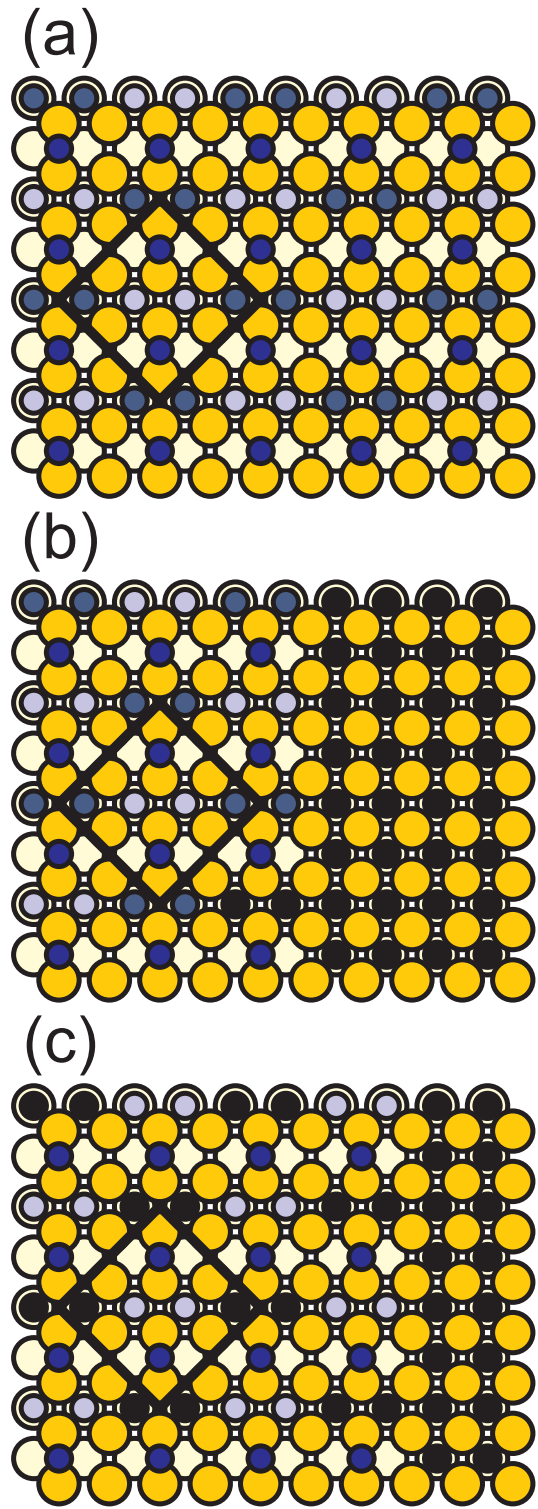}
	\caption{Model of bulk layers for the formation of Mg ferrite and phase separated MgO as concluded from the XPS analysis. (a) magnetite (for comparison). (b) Mg ferrite with small Mg content coexisting with MgO. (c) Mg ferrite with high Mg content (Fe$^{2+}_{\rm{oct}}$ completely substituted by Mg$^{2+}_{\rm{oct}}$) coexisting with MgO.}
	\label{fig:Interdiffusion}
\end{figure}

Here, we assume that Mg$^{2+}$ ions are only located on octahedral sites.
We like to emphasize that our analysis is not sensitive to the coordination of Mg ions.
In principal, these ions may also be situated on tetrahedral sites.
However, this site seems to be quite unrealistic since Mg$^{2+}$ exclusively occupies octahedral sites in both structures MgO (rock salt structure) and ferrites (inverse spinel structure).
As described above only half of the octahedral sites are occupied by cations in the inverse spinel structure.
Therefore, it seems to be obvious that excess Mg$^{2+}$ ions can be placed on the vacant octahedral sites of the inverse spinel structure (cf. Fig. \ref{fig:Interdiffusion}(a)).
These sites, however, are blocked by the Fe$^{3+}_{\rm{tet}}$ ions on adjacent tetrahedral sites.
Thus, the Fe$^{3+}_{\rm{tet}}$ content has to be reduced to gain vacant octahedral sites which are not blocked by Fe$^{3+}_{\rm{tet}}$ ions and can be occupied by Mg$^{2+}$ ions.
In addition, Fe$^{3+}_{\rm{oct}}$ content has also to be decreased due to the 1:1 stoichiometry obtained from the Fe $2p$ analysis.

From these considerations, it is not clear how the sites feasible for Mg$^{2+}$ ions are distributed in the oxidized film.
One possiblity that Fe$^{3+}_{\rm{tet}}$ and Fe$^{3+}_{\rm{oct}}$ free regions are formed only locally and have point-like character.
Another possibility is that these regions free from Fe cations agglomerate and, thus, large parts of MgO are formed.
Thus, the film undergoes a phase separation in coexisting phases of Mg$^{2+}_{\rm{x}}$Fe$^{2+}_{\rm{1-x}}$Fe$^{3+}_{\rm{oct}}$Fe$^{3+}_{\rm{tet}}$O$_4$ ferrite and MgO.
It seems not to be very likely that Mg ferrite and MgO laterally coexist since in this case large parts of the Fe oxide film would be replaced by MgO due to Mg interdiffusion and oxidation.
Therefore it seems to be easier to explain our results be some (at least partial) de-wetting process where Mg ferrite islands are formed on the MgO substrate so that the XPS experiment detects additional Mg $2p$ signal.


Fig. \ref{fig:Interdiffusion} (b) and (c) show schematically the development of the ferrite phase and the MgO phase during the oxidation process while Fig. \ref{fig:Interdiffusion} (a) presents the magnetite structure for reasons of comparison.
In contrast to Fig. \ref{fig:InvSpinel} showing the surface structure bulk layers are presented here.

The stoichiometry after the first oxidation step is presented in Fig. \ref{fig:Interdiffusion} (b).
Large parts of the detected area are covered by MgO while the Mg$^{2+}_{\rm{x}}$Fe$^{2+}_{\rm{1-x}}$Fe$^{3+}_{\rm{oct}}$Fe$^{3+}_{\rm{tet}}$O$_4$ ferrite phase contents only very little Mg.
Approximately one third of the Fe$^{2+}_{\rm{oct}}$ ions are replaced by Mg$^{2+}_{\rm{oct}}$ ions.
After the last oxidation step almost all Fe$^{2+}_{\rm{oct}}$ ions are substituted by Mg$^{2+}_{\rm{oct}}$ ions and almost pure Mg$^{2+}_{\rm{oct}}$Fe$^{3+}_{\rm{oct}}$Fe$^{3+}_{\rm{tet}}$O$_4$ (cf. Fig. \ref{fig:Interdiffusion} (c)).

In addition, our analysis shows that the apparent content of the MgO phase decreases during the formation of Mg ferrite.
Assuming that the detected Mg XPS signal is due to parts of the MgO substrate which is not covered by the ferrite film, one has to conclude that the de-wetting process observed for the ferrite film with low Mg content is reversed and that the ferrite film with high Mg content spreads out.
This effect implies that the surface energy for the ferrite film with the high Mg content is smaller than the surface energy of the film with low Mg content.
Although we did not find any report on the surface energy of Mg ferrite, this result agrees well with literature.
For instance, the surface energy of Ni ferrite is smaller than the surface energy of magnetite \cite{mis77} and substituting Fe$^{2+}$ by Mg$^{2+}$ in wustite decreases the surface energy, too \cite{car90}.

Our LEED studies carried out in addition to the XPS investigations show a $(\sqrt{2}\times\sqrt{2})\rm{R}45^{\circ}$ superstructure independent of the oxidation step.
Since this superstructure is attributed to charge ordering of Fe$^{3+}_{\rm{tet}}$ and Fe$^{2+}_{\rm{oct}}$ in the first subsurface B layer (cf. Fig. \ref{fig:InvSpinel}) \cite{pen05} we conclude that the same charge ordering effect appears for Fe$^{3+}_{\rm{tet}}$ and Mg$^{2+}_{\rm{oct}}$ when the latter has completely substituted Fe$^{2+}_{\rm{oct}}$ (cf. Fig. \ref{fig:Interdiffusion} (c)) as well as for intermediate intermixed structures (cf. Fig. \ref{fig:Interdiffusion} (b)).
This result differs from (1$\times$4) and (1$\times$3) superstructures previously reported for magnetite films after Mg surface seggregation \cite{and97,spi04}.
However, in these studies the magnetite films also deposited on MgO(001) were annealed at much higher temperatures and at different oxygen pressure.
Thus, their formation may be attributed to additional high temperature diffusion effects or vacancy formation while the $(\sqrt{2}\times\sqrt{2})\rm{R}45^{\circ}$ superstructure is stable for moderate oxidation temperatures as used here.


This observation immediately shows that the formation of Fe$_2$O$_3$ in form of maghemite is rather excluded, since maghemite does not exhibit a $(\sqrt{2}\times\sqrt{2})\rm{R}45^{\circ}$ reconstruction. Furthermore, the shape of the satellite structure of the Fe 2p core level spectrum may also change in a different way as observed, as the CTM calculations indicate different shapes for Fe3+ ions in octahedral and tetrahedral coordination, respectively.

Regardless of the underlying model XPS data and CTM calculations clearly show that the alignment of ions in the film is changing during exposure to oxygen.
Although for prolonged oxidation the strong Fe satellites in the Fe $2p$ PE spectra seem to indicate the formation of a defect spinel structure such as $\gamma$-Fe$_2$O$_3$, this is not supported by the LEED results.
In addition, it can be easily derived from Fig. \ref{fig:cluster} that without the presence of Fe$^{2+}$ ions on octahedral sites the correct satellite positions can not be reproduced.
Instead we must attribute the appearance of the satellites to a changing alignment of ions under continued oxidation.
A reason for these changes may be to better balance additional charges incorporated in the film during initial oxidation.
Further investigations are needed at this point to clarify the nature of the structural changes and to determine which of the proposed models is correct.
In particular, site-sensitive methods like X-ray Standing Waves (XSW) may be useful in this regard.


In summary we reported on the characteristics of an iron oxide film prepared by oxidation of a pre-deposited Fe film on MgO(001).
Although the surface structure and the stoichiometry of the film did not change with prolonged oxygen exposure, XPS results indicated changes in the film.
We attributed these changes to the segregation of Mg into the iron oxide film.
The arrangement of Mg and Fe ions on the cation sites in the crystal structure of the film was shown to depend on the duration of the oxygen exposure.
CTM calculations were used to demonstrate that with prolonged oxidation Mg tended to accumulate on Fe$^{2+}_{oct}$ sites.


\begin{acknowledgments}
Financial support by the DFG (KU2321/2-1) is gratefully acknowledged.
\end{acknowledgments}

\end{document}